\documentstyle[12pt]{article}
\input epsf.sty
\newcommand{\tHooft}{{}'t Hooft}
\newcommand{\Ss}{Schwarzschild\ }
\begin{document}
\rightline{Preprint UFIFT-HEP-94-13}
\rightline{hep-ph/9410306}
\vskip.6in

\setcounter{page}{1}

\centerline{\large {\bf Partons and Black
Holes\footnote{Based on lectures given
by L.S. at the ``Theory of Hadrons and Light-front QCD'' workshop
in Zakopane, Poland, August 1994.}
}}
\vskip.3in
\centerline{ Leonard Susskind\footnote{
E-mail: susskind@dormouse.stanford.edu}}
\centerline{ Physics Department, Stanford University}
\centerline{ Stanford, California, 94309}
\vskip.3in
\centerline{ Paul Griffin\footnote{
E-mail: pgriffin@phys.ufl.edu}}
\centerline{ Physics Department, University of Florida}
\centerline{ Gainesville, Florida, 32611}
\vskip.6 in
\centerline{\bf Abstract}
\vskip.1in
A light-front renormalization group analysis
is applied to study matter which falls into massive black holes,
and the related problem of matter with transplankian energies.
One finds that the rate of matter spreading over the black hole's
horizon
unexpectedly saturates the causality bound.  This is related
to the transverse growth behavior of transplankian particles
as their longitudinal momentum increases.
This growth behavior suggests a natural
mechanism to impliment \tHooft's scenario that the
universe is an image of data stored on a $2 + 1$ dimensional
hologram-like projection.

\vskip.1in

\vskip.3in
{\bf 1. Introduction}
\vskip.1in

In  thinking about quantum gravity, and  the problems of
black holes, the relativity community's paradigm
is basically free quantum field theory in a background geometry. But
the more you spend time trying to understand what is happening near
the horizon
of a black hole,  the less free field quantum theory seems to make
sense.
Instead, you find yourself forced back to the kind of thinking that
we're
talking about here at this workshop -- partons, light-fronts,
and simple qualitative
pictures of what happens when particles are at very high
energies.  This is a way of thinking which was popular in
particle physics twenty years ago, and which
has become the way of thinking for the group of people in the
light-front community, a qualitative pictorial way of thinking
about the constituents of matter, in situations where one is forced
to think about cutoffs, and  about what happens
as you move the cutoffs around, uncovering
more degrees of freedom of the system, fluctuating at
higher frequencies and smaller distances.

Our discussion on this will be in very qualitative terms, because the
technical terms describing a complete theory do not yet exist.
What does exist, this way of thinking about black holes,
is to some extent a collaboration (over long distances and
large time scales) between L.S.~and Gerard \tHooft, and combines
the ideas of \tHooft[1], Charles Thorn[2], and L.S.[3], with
the profound insights of Jacob Beckenstein concerning the maximum
entropy of a region of space[4].

Other lecturers at this conference have discussed
a new kind of renormalization that has to be done when doing physics
in the light-front frame.  This new renormalization does not
have to do with large transverse momentum, or small transverse
distance, but with small longitudinal momentum.  Something
happens for small longitudinal momentum which causes new divergences
in light-front Feynman diagrams.
In terms of the Feynman-Bjorken $x$ variable, which denotes the
longitudinal
momentum fraction of the partons or constituents in a high energy
scattering
process, we have to introduce a new cutoff for very small $x$, and
consider the effects of moving around the cutoff.
The Hamiltonian $H$ in light-front gauge is always some object which
is independent of the total longitudinal momentum, divided by the
total longitudinal momentum $\eta$.  This object can be some
complicated
operator which typically, however, adds up to the squared mass of
the system.  It might involve transverse motions and ratios of
longitudinal
momenta, but the total $H$ scales like $1/\lambda$ under Lorentz
boosts
$\eta \rightarrow \lambda \eta$.   We have to think about
a new kind of renormalization when we Lorentz boost systems.  The
degrees of freedom at very low $x$ are ultraviolet in nature, from
the point of view of the light-front Hamiltonian.  (While with
respect to the
spatial variable $x^-$, they are long wavelength excitations,
they are very short wavelength excitations
with respect to the time variable $x^+$.)  And the tool for analyzing
such a rescaling of Hamiltonians is the renormalization group.
The problems of understanding matter as we increase the momentum
indefinitively becomes a problem of locating all the possible
fixed-points of this new renormalization group.  While there
are no known exact fixed-points, there are a variety of
approximate behaviors of systems which will be discussed.

A very simple fixed-point was invented by Einstein and Lorentz,
to describe what happens as one boosts ordinary matter.
They didn't quite get it right, so along came Feynman and Bjorken,
who developed a better behavior for quantum particles.
There is also another fixed-point behavior which occurs in
the context of free string theory.
Finally, we will try to show what the correct behavior
has to be in the quantum theory of gravity, why it has to be the
correct behavior, and what it has to do with black holes, and the
paradoxes of black holes.  We will also briefly explain how this
leads one to believe that the
universe is not what one normally thinks of as three dimensional,
but is, in a way, two dimensional.  In this sense, the universe is an
illusion!

\vskip.3in
{\bf 2. Black Holes and Infalling Matter}
\vskip.1in

Before discussing the fixed-point behaviors, let us consider what all
this
has to do with black holes.  Consider figure~1(a), which is a picture
of a massive black hole in radial coordinates.  On the left of the
diagram
is the singularity, where bad things happen to you, and further
towards the center of the diagram is the horizon.  As
is well known, if you are inside the horizon, all light rays and
all time-like trajectories inevitably lead into the singularity.  The
horizon separates that region of space from the region where you
at least have a chance of getting out.

\begin{figure}[t]
\begin{center}
\leavevmode
\epsfxsize=5truein
\epsffile{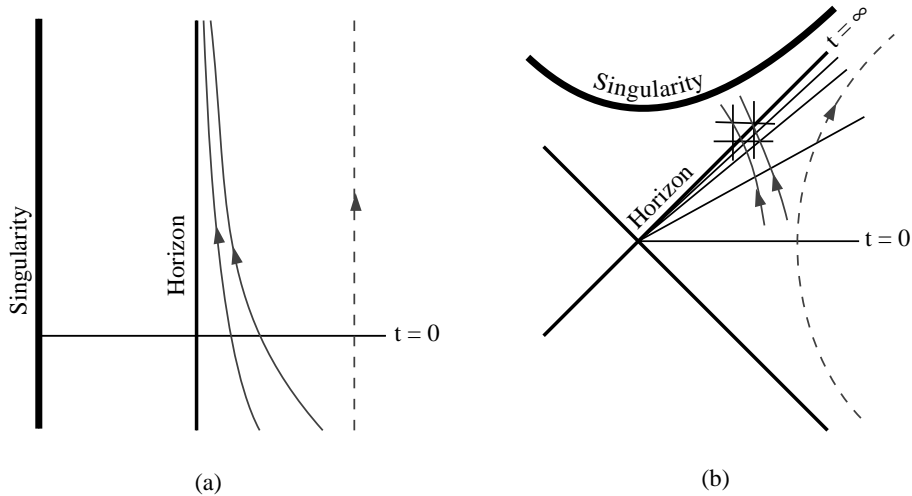}
\end{center}
\caption{(a)~Black hole in radial coordinates.  (b) in Kruskal
coordinates.  Two of the incoming particles in the diagrams
are falling into the black hole.}
\end{figure}
Any matter falling into the black hole, as seen by an
observer outside the black hole in his coordinate frame
and using the \Ss (asymptotic observer) time, will never be seen to
cross the horizon.  (The trajectories of two such observed
particles is shown in figure~1(a).)  As the matter falls in, in the
frame of reference of the static observer, the momentum increases.
It increases for the same reason that the momentum of any object
increases when it falls down, and it's velocity increases, i.e.~it
gets boosted.  It's momentum in fact increases very rapidly with
time -- exponentially rapidly.  Now consider the Kruskal diagram
(figure~1(b)) of exactly the same situation.  Notice that the
singularity
is space-like and not time-like in this
diagram\footnote{That has led some people to ask not where, but when
is
the singularity.  The singularity is in a sense at a place in time.}.
On this diagram light rays move on 45 degree lines, and if you are
in the upper left of the 45 degree Horizon line, you are trapped,
and you have no choice but to fall into the singularity.
If you are in the outer wedge to the right of the Horizon line,
you have a chance of escaping to infinity.  There is a one-to-one
correspondence between the two diagrams.  On the Kruskal diagram,
lines of constant time are lines of constant angle.  The relationship
between one time and another time is a boost, so time is simply
boost angle in this representation.
They get denser and denser as
they approach the horizon line.  The horizon is $t = \infty$, where
$t$ is the time as seen by a stationary observer outside
the black hole.  A fixed static observer moves on the dashed
hyperbola
of fig.~1(b), and she cannot look in or ever see anything behind the
horizon .  Looking back along his light-fronts, all she can ever see
are
particle approaching the horizon, but never passing through it,
because by the time they get to the horizon, she's at her $t =
\infty$.
A second coordinate patch of figure~1(b) is regular from the point
of view of infalling matter.  In terms of ordinary cartesian
coordinates it's metric is given by
\begin{equation}
ds^2\;=\;dT^2-dZ^2-dX^idX^i \ ,
\end{equation}
while in terms of \Ss coordinates the same space-time is
given by
\begin{equation}
ds^2\;=\;{({{dt}\over {4MG}})^2}{\rho}^2-d{\rho}^2-dX^idX^i \ ,
\label{rindler}
\end{equation}
where we have used the coordinate $\rho$ which denotes proper
distance from the horizon instead of the usual radial coordinate $r$.
The Minkowski and Schwarzschild
coordinates are related by
\begin{eqnarray}
Z\; &=& \;\rho\, {\cosh(t/4MG)} \ , \\
T\; &=&\;\rho\, {\sinh(t/4MG)} \ .
\label{minshld}
\end{eqnarray}
There is a big mismatch between the
two coordinate systems as time goes on.  The mismatch is a huge
boost.  That is, the effect of a Schwarzschild time translation on
the Minkowski
coordinates is a boost
\begin{equation}
\Delta \omega \;=\; {{\Delta t}\over{4MG}} \ ,
\label{hyprot}
\end{equation}
and as time goes on, the outside observers coordinate system
and the coordinate system associated with the infalling matter
become infinitely Lorentz boosted relative to each other.
The horizon of the black hole is the surface $t=\infty$.
This point can be made in another way.  As the particle
falls toward the horizon, it is accelerated from the
point of view of a static observer. The momentum
of the particle increases
like
\begin{equation}
P \rightarrow e^{t/4MG} \ .
\label{boost}
\end{equation}
To give a proper account of the particle from the Schwarzschild frame
we must have a description
which is valid at ever increasing momentum.

Black holes, as is well known, evaporate.  They radiate photons,
and other things,  and at the same time, matter may be falling in.
One may imagine throwing in matter at the same rate that the black
hole evaporates
in order to think of a fixed size black hole.  One keeps throwing
in more and more matter -- graduate students, cities, planets,
generally
huge amounts of matter, as long as we do it slowly enough so that
the black hole won't increase its size.  Enormous amounts of matter
accumulate at the horizon.  How does the horizon manage to hold all
that material?  How does it manage to squeeze that much in?
The answer, according to the usual picture, is that the matter
Lorentz contracts.  Because the
momentum given by eqn.~(\ref{boost}) becomes so large so
quickly as the objects fall in, they "pancake" and get flattened.
They get so incredibly flattened, that you can apparently
put any amount of matter on the horizon.  The Hawking radiation
apparently has nothing to do with the incoming matter -- it is
produced
by curvature at the horizon, and it not the matter that is falling
in,
boiling, and coming back out.
While this decoupling between infalling matter and Hawking radiation
is
almost certainly wrong,  the point of this discussion is to
make one realize that in the study of matter falling onto the
horizon, one is encountering huge boosts.  How long do we have
to follow this matter?  For the purposes of this discussion, we
have to follow it for a time comparable to the lifetime of the black
hole.
The lifetime of a black hole is proportional to $M^3$, so if we
wait long enough, we will find that the momentum of the infalling
matter
has grown to
\begin{equation}
P_{\rm lifetime} = e^{M^2 G} \ .
\label{bigboost}
\end{equation}
We have to track the matter until it's momentum is an
exponential of the black hole's mass squared in Planck units.  A huge
amount of momentum is involved even if the mass of the black
hole is a hundred or so Planck masses.  This energy
is vastly bigger then the entire energy in the universe!
Therefore, in order to understand and study the properties
matter falling into a black hole, we have to understand the
boost properties of matter far beyond the Planck scale.

\vskip.3in
{\bf 3. Boost Properties of Matter}
\vskip.1in

\begin{figure}[t]
\begin{center}
\leavevmode
\epsfxsize=4truein
\epsffile{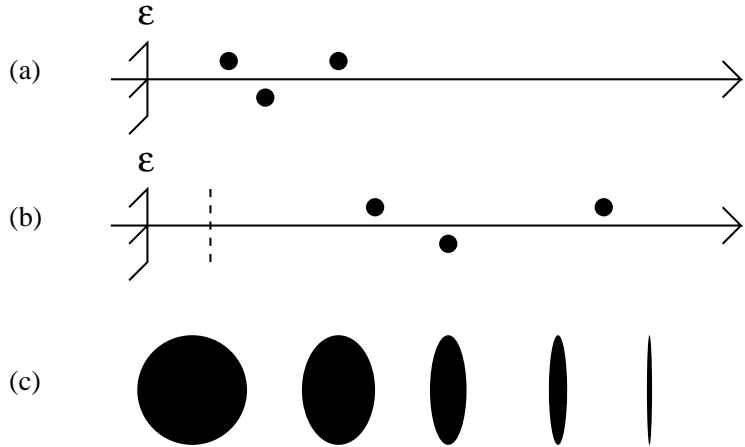}
\end{center}
\caption{The Einstein - Lorentz fixed point under boosts.
(a) Initial Parton distribution on the longitudinal momentum axis,
$\epsilon$ is a momentum cutoff.
(b) Parton distribution after a boost which doubles constituent
momentum.
(c) Pictures of the matter undergoing greater and greater boosts.
A pancake picture of the matter emerges. }
\end{figure}
The boost properties of matter are described by the fixed points of a
new kind
of renormalization group.  Let us consider some of the qualitative
aspects of the speculations that have already been put forward about
these
fixed points\footnote{They were not called fixed points when they
were
originally worked out, but now it is worthwhile to think of them this
way.}
Consider the longitudinal momentum of a particle.  We need to cut off
very small longitudinal momentum with a cutoff $\epsilon$, because
that region corresponds to very ultraviolet behavior in the
light-front
frame[5].  Physically, we imagine that we have a set of
detectors or apparatuses that are sensitive to
frequencies up to some maximum of order $M^2{\epsilon}^{-1}$. An
appropriate description should be possible with longitudinal momenta
cutoff at $p^+=\epsilon$. It is important to keep in mind this
connection
between the cutoff procedure and an
apparatus.

\vskip.2in
{A. The Einstein Lorentz Fixed Point}
\vskip.1in
In the picture of matter developed by Einstein and Lorentz, the
particles consist of only a few `valence' partons, shown on the
longitudinal
momentum axis in fig.~2(a).
When we boost the matter
by some factor, each parton constituent of the matter gets boosted by
the same factor.  In this E-L picture, transverse position of the
partons
doesn't change.  So when the momentum of the matter is doubled, the
longitudinal momentum of the partons which make of the matter is
doubled.
One has also doubled the differences between the momentum of the
partons, and doubled the fluctuation in momentum.  So by the
uncertainty principle, you have contracted everything by the same
factor of two.  The E-L fixed point is described by the
behavior of matter shown in figure 2(b) for different values of the
boosts.  When the particle is at rest, it looks like a sphere.
When the particle is moving, it looks like an ellipsoid.
As it moves faster, it looks like an even flatter ellipsoid.
Send it to extremely high energy, and it looks like a
flat pancake, with infinitely thin dimensions.  That is the
E-L fixed point.  If
the cutoff $\epsilon$ is sufficiently small the probability to find a
parton with $p^+ <\epsilon$
is vanishingly small. Boosting the system is trivial, as each parton
shifts to its boosted position.
Transverse sizes are invariant and longitudinal sizes contract.
Most likely no real 3+1 dimensional quantum field theory works this
way.

\vskip.2in
{B. The Feynman Bjorken Fixed Point}
\vskip.1in
\begin{figure}[t]
\begin{center}
\leavevmode
\epsfxsize=4.5truein
\epsffile{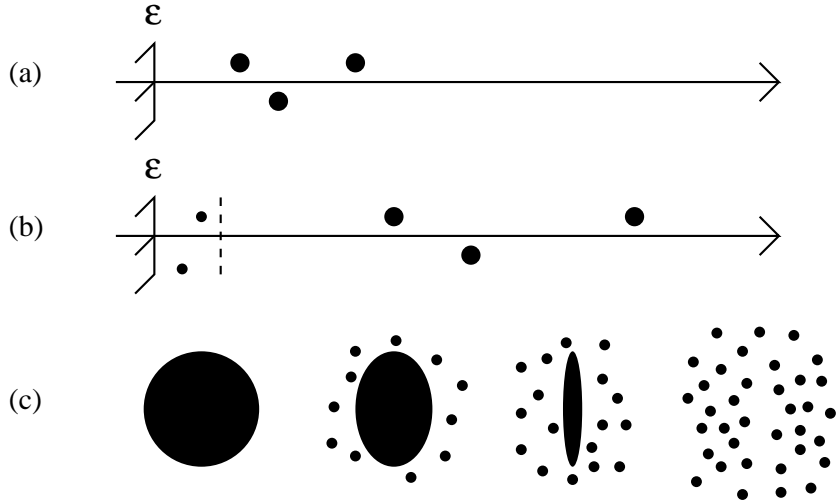}
\end{center}
\caption{The Feynman - Bjorken fixed point under boosts.
(a) Initial valence parton distribution on the longitudinal
momentum axis.
(b) Parton distribution after a boost which doubles constituent
momentum.
Wee partons appear from the vacuum effects below the old (now
boosted) cutoff
and above $\epsilon$.
(c) Pictures of the matter undergoing greater and greater boosts.
A pancake picture of the matter emerges, with an unboosted and
transversally
widening cloud of wee partons.}
\end{figure}
The next fixed point is the Feynman-Bjorken (F-B) fixed point[6,7].
It is a lot like the E-L fixed point with respect to the valence
partons.  These partons
carry the information of the matter that tells it that it's
a proton and not a neutron -- isospin, charge, etc, and there
is only a few of them.  When you double the momentum of the matter,
these valance partons also double their momentum.  If we drew the
same sequence of boosts for the valence partons, we would expect
the same E-L pancake behavior.  But according to Feynman
and Bjorken, maybe there are degrees of freedom behind the
$\epsilon$ momentum cutoff, which we don't care about because
they are fluctuating too rapidly for us to see.  Now we boost
everybody, and some degrees of freedom which might have been behind
this cutoff now come out into direct
view\footnote{The cutoff is kept in the
same place under boosts.  Note that we could also keep the momentum
fixed and decrease the cutoff.},
where our physics should be able to see it.
The very high frequency fluctuations become visible
when we further boost the system because they slow down.
In this model of hardons, the the number of partons per unit
$p^+$ behaves like
\begin{equation}
{dn\over{dp^+}} \sim {1\over{p^+}}
\label{partdis}
\end{equation}
So, in the F-B fixed point picture, every time we
boost matter, we pull some more partons out of the
vacuum.  In other words the boost operator must contain an extra
term which acts as a source of partons of low $p^+$. These are the
partons that Feynman called the `wee' partons.
The wee partons create interesting
anomalies in the behavior of
matter under boosts. For example, because they always carry low
longitudinal momentum they
contribute a cloud which does not Lorentz contract.
Also, while Feynman and Bjorken did not say this,
one believes that they tend to spread out a little in the transverse
direction, as the system is boosted.
For example, in simple Regge approximations their average distance
from the center of mass satisfies[8]
\begin{equation}
R^2_{\perp}({\rm wee}) \sim log{{p^+({\rm tot})}\over {\epsilon}}
\label{herewee}
\end{equation}
So this improved F-B fixed point
is the following.  At rest, the matter is a sphere with no wee
partons.  Boost it and the valence stuff behaves as in the
E-L picture and pancakes, but the wee partons form a cloud around
it which does not Lorentz pancake, but may spread out a little
bit transversely.  This is shown in figure~3.

\vskip.2in
{C. The Kogut-Susskind-Lipatov-Alterelli-Parisi Fixed Point}
\vskip.1in
The next fixed point to consider is the
Kogut-Susskind-Lipatov-Alterelli-Parisi [9] (KSLAP)
one, which attempts to treat another aspect of this problem.
If you look at things with better and better transverse
resolution,
you begin to see matter consisting of smaller and smaller parts.
At low momentum, a meson would look like two partons, but if you
increase the transverse momentum, you might see ``new'' wee partons,
but
also if you look at the original valence partons more carefully,
you might also see that they consist of partons within partons.
For example, a quark might spend part of its time being a quark and
a gluon.  A gluon spends part of its time being a pair of quarks.
This is shown schematically in figure 4.
\begin{figure}[t]
\begin{center}
\leavevmode
\epsfxsize=4truein
\epsffile{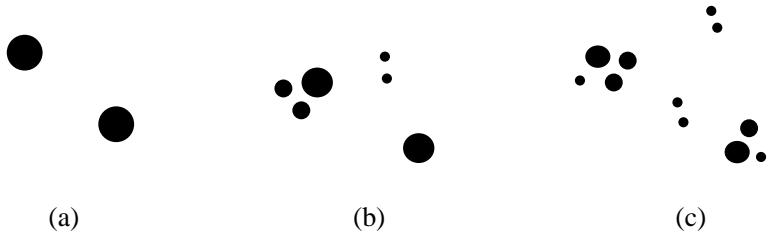}
\end{center}
\caption{The behavior of mater with respect to boosts in the
KSLAP fixed point picture.  As the resolution increases
(fig.~(a) $\rightarrow$ fig.~(c)), you see
both new wee partons, and valence partons that spend time
as multi-parton configurations.}
\end{figure}
As we boost the system so that the parton has a much larger
$p^+$, the transverse phase space for it to split is much bigger.
If the theory has transverse divergences then that
probability will become large as the system is boosted. Eventually
the parton will be replaced by two or more partons closely
spaced in transverse space. The effect continues as the system is
boosted so that the partons reveal transverse fine structure within
structure ad infinitum.
One can do a great deal of physics in the study of deep inelastic
scale violations by thinking naively in this
physical way.  In fact, one important consequence for
high energy scattering is the existence of processes
involving large transverse momentum jets.

\vskip.2in
{D. The String Theory Fixed Point}
\vskip.1in
Let's think of strings as being made up of balls and springs.
(Forget
for the moment any fancy mathematics.)  What happens as you add more
balls and springs, in an effort to try to define a string?  You
add more and more normal modes of higher and higher frequency.
To understand the low frequency aspects of the string, you can simply
think of it as a couple of balls and springs. (From now on,
low momentum means Planck scale!)
If you want to
get all the high frequency modes right, you have to add a lot of
balls and springs. At some relatively low Planckian momentum,
let's assume a particle
consists of two partons -- two balls and a spring.  This is
just our longitudinal momentum cutoff because we don't see
very high frequencies.  As we probe the string at higher
and higher momenta, we need to add more and more balls and
springs, and it seems to the observer that the original
constituents are breaking up into more
constituents, simply of the same low momentum, with the number
of total constituents being the total momentum of the particle.
The normal modes of the string fluctuate in the transverse direction,
yet you see more of them as you probe higher and higher
light-front energy.  In string theory, transverse momentum never
get big, and we are also seeing here that it also says soft
longitudinally.  So, instead of boosting the particle by increasing
the momentum of the constituents, you just increase the number
of constituents.
Notice that since you never make large momentum constituents,
there is no reason to think that this object Lorentz contracts.
Every parton is wee inside the string.  Charles Thorn
calls this just the fact that the string is made of wee
partons[2].  Mathematically, this is related to
`conformal invariance' of the string.  The pattern of
growth in transverse position is shown in figure 5.
\begin{figure}[t]
\begin{center}
\leavevmode
\epsfxsize=4truein
\epsffile{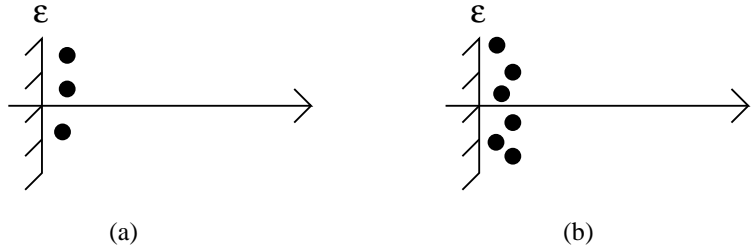}
\end{center}
\caption{The string theory fixed-point.  As the momentum increases
(fig.~(a) $\rightarrow$ fig.~(b)), more partons appear.
All partons are wee, and there number is
proportional to the transverse radius squared. The momentum
of the string grows like $P = 2^N$, where $N$ is the number of
partons.}
\end{figure}
It's called ``branching diffusion''.  Each time you double the
momentum,
each constituent bifurcates.  It's like the evolution of a petri dish
full of single organisms, except in free string theory those single
organisms don't push each other out of the way.  A low energy string
might consist of just two balls, with a size and separation
characteristic of the string scale.  Double the momentum and each
ball
bifurcates into two balls with some random orientation in the
transverse direction, each one of which is about the same size
as the original one, and the distance between them is about the
same characteristic string scale.  Eventually, for a large amount of
bifurcations, the total momentum is
\begin{equation}
P^+ ({\rm tot})  = \epsilon\, 2^n \ ,
\end{equation}
where n
is the number of partons. The mean square
radius of this object is defined as follows: each time it
bifurcates, focus onto one of the bifurcation products
randomly, and go to the next one, etc.  As you do so,
you will be forming a random walk in the transverse plane
in coordinate space.  So the mean square size of the object
will grow like
\begin{equation}
R^2 = \ell_s^2\, n = \ell_s^2 \, \log (P^+ / \epsilon ) \ ,
\label{rstring}
\end{equation}
where $\ell_s$ is some characteristic string length scale that
is of order of, or greater than, the Planck
length\footnote{
Note that eqn.~(\ref{rstring}) is a Reggie pole formula,
and explains why string theory has Reggie pole behavior.
}.
The precise relation is
\begin{equation}
\ell_{\rm Planck} = g\,  \ell_s
\end{equation}
where the string coupling $g$ is much less than one for weakly
coupled strings.  We will basically assume that strings are
weakly coupled in this lecture.

Let's see that the mean square radius of an object does indeed
increase
in this way in string theory, and how it's connected to this new kind
of
renormalization group thinking.  A string in the transverse plane is
parameterized as
\begin{equation}
X^{\perp} (\sigma ) = X^{\perp}_{\rm cm} +\sum_{l}
\frac{a^{\dagger}(l)}{\sqrt{l}} e^{il (\sigma +\tau)} + \cdots
\label{xperp}
\end{equation}
where $0< \sigma< 2\pi$ labels the position along the closed
string, for fixed light-front time $\tau$, and
$X^{\perp}_{\rm cm}$ labels the center of mass of the string.
The rest of the expansion in eqn.~({xperp}) is a bunch of normal mode
oscillations.  Each of these oscillations has higher and higher
frequencies
in light-front time.  The mean squared radius of the probability
distribution of a point on the string, where we have subtracted the
center of mass contribution, is given by the formula
\begin{equation}
R^2 = \sum_l \frac
{\langle a^{\dagger} (l) a(l) + a(l) a^{\dagger} \rangle}
{l}
\sim \sum_l 1/l \rightarrow \log \infty \ .
\end{equation}
\begin{figure}[t]
\begin{center}
\leavevmode
\epsfxsize=5truein
\epsffile{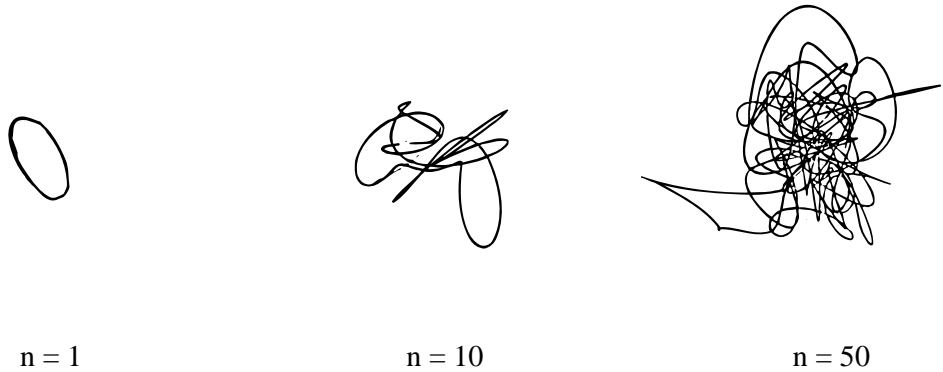}
\end{center}
\caption{Snapshots of a string's transverse
extent as more and more modes $n$ are included.  Ultimately
it becomes space filling as $n$ goes to infinity.
}
\end{figure}
Therefore, it is a complete lie that strings are little objects
about as big as the Planck scale
\footnote{
This is what made string theory
fail as a theory of hadrons, since hadrons are not infinitely
big.
}.
However, what is wrong with this picture is that the infinity
is coming from the incredibly high modes of oscillation that we
can't see.  They are behind the low $k^+$ cutoff in this sense.
How many modes should we account for?  The number of modes
we should account for is comparable to the momentum, because
the time dilation factor of a fast object is comparable to its
momentum.  Therefore we find that
\begin{equation}
R^2 \rightarrow \log P \ ,
\end{equation}
and that is the same rate of growth given by the bifurcation
for the string fixed point.  The total
length of string grows as $P$ itself, so the momentum per unit
length is approximately constant.  Figure 6 shows some snapshots
of a string as you include higher and higher normal modes[10].
These are Monte Carlo samplings of string configurations drawn
from the wave functions of strings with given numbers of included
modes.  We can think of these pictures in several different ways.
We can think of them as strings where you include more and more
modes.  Or, we can think of this as strings at even increasing
momentum.  Or, we can think of these as the description of a string
as falls towards a black hole.

\vskip.3in
{\bf 4. The String Picture of High Energy Scattering and Matter
Falling
into Black Holes}
\vskip.1in
Now let us discuss high energy scattering amplitudes, and let's
see why string theory is quantum gravity and not quantum
`something else'.  Take two particles scattering off each other,
at first in a weakly coupled approximation, and we won't
worry about unitarity for the moment.  One particle has momentum $P$
and
the other momentum $Q$ in string units.
Following the string theory picture of the
constituents, the first contains $P$ partons, and the second
$Q$ partons.   It turns out to be easier to think about the
interaction
in the center of mass frame of the two particles, although
we still have the picture of their constituents in the infinite
momentum frame.  We assume that the scattering amplitude is the
sum of parton -- parton scattering amplitudes, and will think of this
later on as gravitational scattering, but always gravitational
scattering
between low momentum constituents.  The energy of the constituents
is not increasing, and the scattering amplitude of the constituents
is independent of the momentum $P$ and $Q$.  We therefore expect the
forward
coherent scattering amplitude will go like $S = P * Q$, where $S$ is
the
Mandelstam $S$ variable.  The energy of the system is roughly $E^2 =
P * Q$,
so $S$ goes like $E^2$.  In Reggie theory,
a pole with intercept corresponding to spin $s$ exchange contributes
as
$S = E^s$, so our result is exactly the rate of growth of the
scattering amplitude for a spin two exchange.
That is the graviton trajectory, so if this theory can be made to
fly, it will
automatically have a graviton Reggie trajectory.

Now this is far from unitary as it stands, since the particle grows
only
logarithmically in transverse size as you increase the number of
partons,
but we are saying that the forward cross section
(proportional to $S$ by the optical theorem) grows linearly
as you increase the number of partons.  In a unitary theory, a cross
section
cannot be bigger than the particles which are scattering,
they can't be bigger than the geometric size of the object.  We will
resolve this paradox in section~5.

In the standard picture of the relativists, who think about free
particle
quantum field theory, the size of a particle is longitudinally
boost invariant, so as it falls toward a black hole,
a point particle remains a point-like object which
asymptotically approaches the horizon.  What does string theory say?
As you approach the horizon, $R^2 = \ell_s^2 \log P$,
from eqn.~(\ref{rstring}).   We also know
from eqn.~(\ref{boost}) that as a particle falls towards the horizon,
its momentum increases exponentially rapidly, so
the radius increases rather substantially as
\begin{equation}
R^2 = \frac{\ell_s^2\ t}{4MG}  \ .
\end{equation}
(The $1/4MG$ factor is just the time dilation factor as
seen from spatial infinity.  An observer near the horizon would
not see this factor.)  This growth is characteristic of diffusion.
Now recall that the radius of a black hole is $R_{\rm schw} = 2 M G
$,
so we can calculate how long it will take for the string
to spread across the entire black hole.  We find that
\begin{equation}
t_{\rm speading} = \frac{G^3 M^3}{\ell_s^2} \ .
\end{equation}
This time to spread across the entire black hole is small compared
to $G^3 M^3/ \ell_{\rm Planck}^2$,  which in turn is the time it
takes to
evaporate the black hole altogether.  With this very
modest $\log {}$ growth in the string size, there is enough time
for the string to spread across the entire horizon.

This is a particularly strange result for the following reason.
Somebody falling with the string does not get to see these increasing
number of normal modes\footnote{
Consider throwing a hummingbird into a black hole.
As it approaches the black hole,
the motion of its wings slow down, because it is being time dilated,
or alternatively, because it's increasing its momentum.  So we see
it going from just a little body with nothing else there,
to a thing with some wings.  What does a hummingbird think about
all this?  Not much, because it doesn't know it has wings
since it has never been time dilated relative to itself,
and it can't see its wings as it falls into the horizon either.
}
So the matter which falls in doesn't know that it's spreading, or
``melting", but an outside observer can see this quite clearly.
As long as the observer falling into the horizon cannot send
out a message to the
asymptotic observer, there will be no paradox as to the
description of its extent along the horizon.

\vskip.3in
{\bf 5. Unitarity and Beckenstein}
\vskip.1in

Consider two particles, which are balls of wound up string,
whose radius grows only logarithmically in energy as shown in the
previous section.  We seem to have calculated that their
elastic cross section grows as a power law in the energy,
which is a violation of unitarity.  According to
Froissart[11], the cross section of such an event should
never grow faster than a logarithm of the energy
squared\footnote{
The Froissart bound is $\sigma({\rm total}) < C \log^2 E$,
as the incident particle energy $E\rightarrow \infty$.  The
rigorous derivation of this bound is based on unitarity and
the domain of convergence of the partial wave series for the
imaginary part of the amplitude.  If the partial wave amplitudes
are not normalizable due to long range interactions, then the
derivation
can be invalidated.
}
, so there
must be huge shadowing corrections if in the lowest order,
we seem to find the power law.  At least in ordinary physics
that would be the case as we try to stack up more
and more matter onto the incident bundles of matter.
The other alternative is that the matter, because of
strong interactions, pushes itself away, so that the
geometrical radius grows faster than a log.  The rate
of growth that was described in section 3D is what
happens when you have a bunch of free micro-organisms
growing on a petri dish, which don't push each other
out of the way. At some point, you get a huge
density of them in the center, because the radius is only
growing logarithmically, but the number is growing linearly.
They are going to get incredibly dense.  When they get this dense,
no matter how small the coupling constant, they will start
to interact.  When they start to interact, we propose that
instead of shadowing being the solution to the unitarity problem, the
actual growth is faster than we had estimated.  If they
push each other out of the way in the transverse plane,
then the area will grow proportional to the number of
constituents.  The number of constituents is proportional to
the momentum, since every parton is wee, and therefore the
area $A$ of this object will grow like the momentum,
\begin{equation}
A \propto \frac{P^+}{\epsilon} \ .
\label{areamomone}
\end{equation}
This is deeply connected to another fact -- Beckenstein's
observation[4] that the entropy $S$ of any system can never be bigger
than a quarter of the area,
\begin{equation}
S_{\rm max} =  A /4 \ ,
\end{equation}
in Planck units.  The connection is that Beckenstein gives us
a bound on the number of degrees of freedom per unit area.
You cannot pack stuff more solidly per unit area than one bit
of information per Planck area.  And therefore, the bits
which form a particle cannot stack up to more than one per
unit area, and they are forced to spread out.  This effectively
determines the proportiality constant in eqn.~(\ref{areamomone}),
\begin{equation}
A \sim \ell_{\rm planck}^2 \frac{P^+}{\epsilon} \ .
\label{areamomtwo}
\end{equation}
The area law of eqn.~(\ref{areamomone}) can, in fact, be proved
by considering the following experiment.  Take a target,
a piece of foil for example, and consider what happens
when you smash a billion Planck energy particle into it.
When a particle get such a large momentum,
much bigger than the Planck scale, we can use classical physics
to describe what happens, since energy being large in gravity
theory is like charge being large in QED.
When the charge gets large, the electro-magnetic field becomes
classical.  When the momentum of a particle becomes sufficiently
large, the gravitational shock wave that travels with it
becomes very classical.
The particle will blast out a hole from the target,
and an operational measure of the size of the particle
is the size of the hole.
Let's go now to the center
of energy frame.  Let the incoming particle have energy
$E_{\rm cm}$.  That energy stored in a system at rest would have
a \Ss radius.  If these two particles have an
impact parameter smaller than this radius, then they
will form a black hole.  That is a significant collision,
a process in which a large number of secondary particles
come out -- the Hawking radiation from the new black hole.
Particle physicists would just say that a big
collision happened, and the fireball heated up, and it  was a
very inelastic collision.  Therefore, from classical physics,
we know that the maximum impact parameter $b_{\rm max}$ at
which a collision takes place is
\begin{equation}
b_{\rm max} \simeq G E_{\rm cm} = G \sqrt{S} \ ,
\end{equation}
where $S$ is the Mandelstam variable.
Going back into the laboratory frame, $S = E_{T} P_{\rm lab}$,
where $P_{\rm lab}$ is the momentum of the incomming particle.
Therefore the area of the particle, effectively given by
$A = \pi b_{\rm max}^2$, grows like
\begin{equation}
A \approx \ell_{\rm planck}^4 E_T P_{\rm lab} \ .
\label{areamomthree}
\end{equation}
Furthermore, by
comparing (\ref{areamomthree}) with
(\ref{areamomone}) we find that the role of the parameter
$\epsilon$ is played by
\begin{equation}
\epsilon= \frac{1}{ E_T \ell_{\rm planck}^2 } \ .
\end{equation}
It is not surprising that the ability of a target apparatus
to detect high frequencies
should be limited, through the uncertainty principle, by its energy.

What does that mean for a particle falling into a black hole?
This implies that it's area grows much faster than the previous
string theory based argument.  Instead of $R^2$ growing like
asymptotic time $t$, we now find
\begin{equation}
R^2 \sim P \sim e^{t/4MG} \ .
\end{equation}
This is exactly the maximum rate of growth
that a perturbation near the horizon of
a black hole can spread out by causality.
So we are finding this behavior by studying the longitudinal
boost properties of matter,
and find that it spreads over the horizon\footnote{
The spreading is of course bounded by the area of the horizon.
We cannot study this problem fully because once particle's transverse
size gets to be of order the horizon, our flat space string theory
description of its behavior breaks down.  No one can follow the
evolution of the string past the point where it begins to feel the
curvature.
}.

\vskip.3in
{\bf 6. Counting the Degrees of Freedom of Spacetime}
\vskip.1in
There is an odd and confusing point that seems to have become
clear after conversations between myself (L.S.) and Gerard \tHooft\
this summer, which is that the world that we see is in a certain
sense
not three dimensional, but is two dimensional.  In {\tHooft}'s words,
the world is a kind of illusion, analogous to a hologram.  Its
information
can be stored in two dimensions, but nevertheless can be looked at
from different angles.  The transformations between the description
of the information as seen from different angles are
called the angular conditions by light-front enthusiasts.  These
difficult transformations that rotate the light-front frame are
those which \tHooft\ thinks of as looking at a Hologram from
different
angles.  Let us try to see more
precisely what the connections are between the holographic theory
and the light-front pictures that light-front community has been
developing for the study of QCD.

First of all, why does one think that the world is two dimensional
instead of three dimensional?  The idea goes back to Beckenstein,
who says that the entropy of a black hole counts the number
of degrees of freedom in a region of space-time, and that this
entropy is much less what you might ordinarily expect.
In ordinary three dimensional (plus time) physics, we imagine space
being
filled with degrees of freedom.  Many people would like to propose
that space is either discretized or somehow regulated so that the
number of
degrees of freedom per unit volume is no larger than one
per Planck volume  $\ell_{\rm planck}^3$, which is nice because
then they imagine that their field theories can be finite.
For the moment, let's use some kind of lattice structure like this
simply to count the number of degree of freedom.
If the degrees of freedom were just bits of binary information,
then the number of states such a system has would be
$2^V$, where $V$ is the volume measured in Planck units.
The number of degrees of freedom would clearly be proportional to the
volume,
and the maximum entropy would be
\begin{equation}
S = V \log 2 \ .
\label{entropy}
\end{equation}
That's a plausible counting rule for all ordinary low energy physics.
The basic concept, due to Beckenstein, which changes this counting
rule is the following.  According to Beckenstein, the entropy
of a black hole is proportional to its area.  And we know that
most of the states in the 3-D lattice picture of space-time are
high energy.  We should exclude states which have so much
energy inside the volume $V$ that they form a black hole
bigger than this volume.  Clearly, we are over counting states
if we're including states which have so much energy that
the energy is bigger than a possible black hole of that size.
So consider the following question --  can we find in the volume
$V$ a small region of space which has an energy which
is smaller than that of a black hole of volume $V$, but which has
entropy larger than the black hole?
Suppose we found this object in the center of the cubic volume.
Now create a black hole around it by simply imploding a shell of
matter
so that this energy plus the energy in the small region makes a black
hole
of the appropriate size inside $V$.
Then, we have just created a black hole around
the region, and while originally the entropy of the region was bigger
than the entropy of a black hole surrounding $V$, now that
we have created the black hole, the entropy is smaller than
the entropy which was in the region by assumption.
This system beats the second law of thermodynamics.  So
if we believe the second law, as we do, then we cannot have
the number of degrees of freedom in the region $V$ which is larger
than its surface area in Planck units.

\tHooft\ says that what this means, is that in fact we have
an entire description of everything inside the region if we have
a sufficiently detailed map, with Planckian resolution, of everything
on the surface of the volume.   If we could see and watch what was
going on on the surface, with sufficient detail and precision,
not an infinite precision, but enough to see Planckian cells, with a
binary decision in them, then we have all the information
inside the region.
Let's take the volume $V$ and the sphere encompassing it to be
very large, and stand near the sphere, and look at the world
through the sphere.  Then our view through the sphere is approximated
by a plane, or window, and everything behind that window can really
be
coded by pixels on the window of one pixel per Plank area,
with a pixel being a quantum mechanical binary bit.  This is
\tHooft's two dimensional world, shown in
figure~7.
\begin{figure}[t]
\begin{center}
\leavevmode
\epsfxsize=2truein
\epsffile{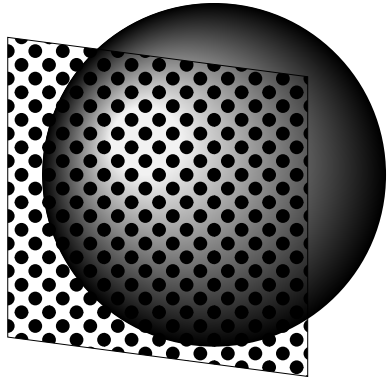}
\end{center}
\caption{\tHooft's holographic representation of the world:  the
pixels on the window at the edge of the sphere encode all the
information on the surface of the sphere.  The surface of the
sphere, by an entropy argument in the text, in turn
encodes all of the information inside it.}
\end{figure}

\vskip.3in
{\bf 7. Encoding Three Dimensional Information into Bits on the
Screen}
\vskip.1in
Let us consider a few examples of how this information might be
encoded
onto the screen of figure 7.  The picture of the world
is that it is made of tiny information `bits', which are
binary in nature, and
everything in the world in made up of the same bits.
Just to get started, let us first consider static configurations.
We want to code the information of the static configuration on a
screen.
Consider the following rule.  Take a light ray associated
with a parton, and simply pass it through
the screen perpendicularly, as in figure 8.
\begin{figure}[t]
\begin{center}
\leavevmode
\epsfysize=2truein
\epsfbox[125 370 625 700]{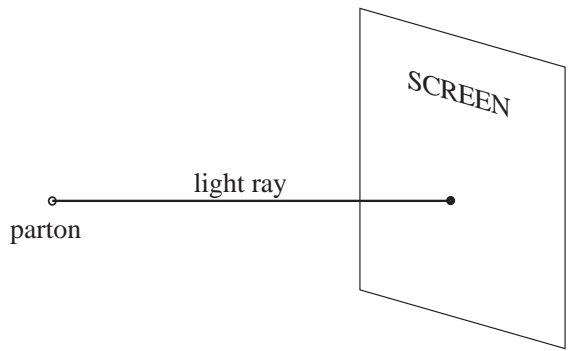}
\end{center}
\caption{A single parton projected onto the screen by
light ray.}
\end{figure}
Imagine that the screen in
composed of a bunch of `light tubes', and that the observer sees
the universe by looking at the tubes.
We will argue that you can never store more information
along a column perpendicular to the screen than one bit per unit
area.   The argument basically constructs a mapping between
bits on the screen and degrees of freedom in space away from the
screen, using the rule of figure 9,
so the final product here will be a mapping or encoding
of three dimensional information on the two dimensional screen.
A column density is a density per unit area where you take everything
behind the screen and you simply make a long straight thin column
of width $\ell_{\rm planck}^2$.  The amount of matter inside is the
column density.  There is separate information along the column,
and we want to map that information to different bits on the screen
without
exceeding one bit of information per Planck area.  Let's try to store
first the maximum amount of information in the 3D region next to the
screen.
According to the previous section, to do this we make a black hole.
\begin{figure}[t]
\begin{center}
\leavevmode
\epsfysize=2truein
\epsfbox[-150 320 850 700]{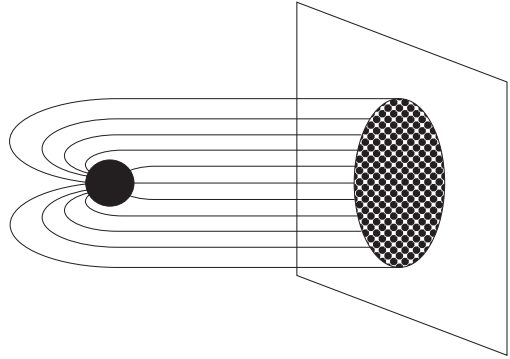}
\end{center}
\caption{A black hole projected onto the screen.}
\end{figure}
The black hole has one bit per area of information on the
horizon.  Now let's use our rule, and take a light ray from
every point on the horizon and project it onto the screen.
The picture of how to do this is shown in figure~8.  In thinking
about the light ray lines, it is always easier to think of the
light rays as coming from the screen and following geodesics
down to the horizon of the black hole.  The black hole forms a
disk on the screen.  What can we say about the mapping from the
horizon to the screen?  The mapping should not take an area on the
horizon and project it to a smaller area on the screen.
Otherwise, we would be packing more than one bit per Planck
area of information onto the screen.  The focusing theorem[12]
guarantees that this won't happen   It says that if you follow a
bundle of light rays, the second derivative of the
area of the bundle  with respect to the affine parameter
is always negative.  The more matter that is in the
black hole, the more negative it is.  In other words,
matter tends to pull light rays together.  Since these
light rays start out parallel at the screen,  the area
of a bundle of light rays always decreases as you move away from the
screen.   Note that at the pole of the black hole closest to the
screen, the mapping is approximately area preserving, so you know you
have to pack one bit per unit area on the screen.
So far, it is not surprising that we have not succeeded in filling
the screen up with a higher amount of information than one
pixel per Planck area.  So consider putting some more stuff in
behind the black hole, for example another black hole.
What do the light rays do for this configuration?
Gravitational lensing occurs, and light rays don't pass through black
holes,
they go around black holes.   What you see is an Einstein ring
around the disk from the closer black hole.   The rays which make
the ring can be traced back to the second black hole.  No information
is lost,
even though some of the rays you might
trace from the second black hole fall into the first black hole.
Put another way, there is always some light ray which joins a
bit on the horizon of the second black hole to the screen.
There is no shadowing in gravitation.  You can't hide behind a black
hole.  If you try to hide behind one, then someone can see your
Einstein ring.  The more bits you put behind the first black hole,
the further out the rings will appear.

Supposing we are interested in a black hole slowly being eclipsed
by a closer one.  The the screen will see a sequence of patterns
as shown in fig.~9.
\begin{figure}[t]
\begin{center}
\leavevmode
\epsfysize=3truein
\epsfbox[-150 320 850 800]{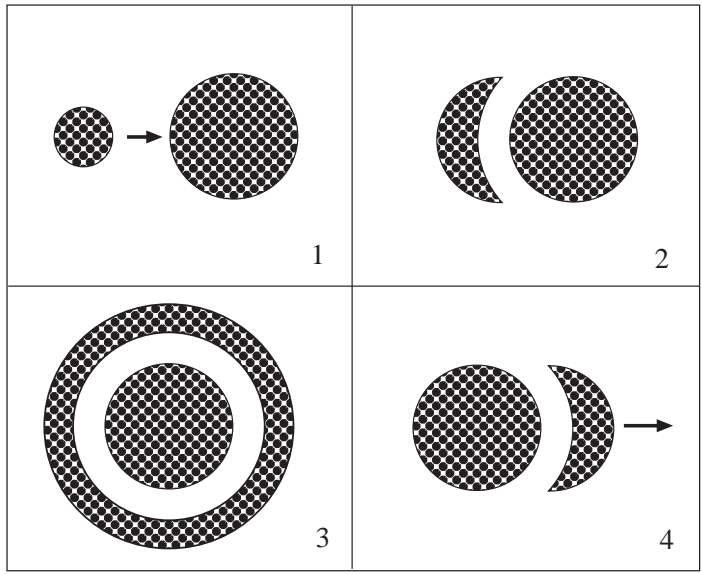}
\end{center}
\caption{One black hole passes behind another, producing
an Einstein ring of information, similar to gravitational lensing,
on the screen.}
\end{figure}
Initially, you see two disks of the two black holes, as they get
closer, assuming they do not collide,  the image from the
one behind starts to get deformed,  then at `total eclipse' you
get an Einstein ring, and finally a two disks as they draw
apart.
If they collide and merge, they will just form another black
hole, and the density of information will still not be bigger
than one bit per Planck unit on the screen.

This picture makes you think that information is behaving like a
fluid which is repulsive and incompressible.  Somehow
gravity must have some component to it which is repulsive,
always preventing you from saturating the screen
with arbitrarily high information density.  In this sense,
bits have elbows.

\vskip.3in
{\bf 8. Encoding Information of Moving Particles}
\vskip.1in

Now let's go a little further.  We have developed a
description which is necessarily a little imprecise
because we want to be doing quantum mechanics, and the
above was classical.  The next step is to think about matter which
is moving around and so {\it what we have to map is not points
of space to the screen, but points of space-time to a 2 + 1
dimensional space-time}.  We do this in the same way -- take a
light ray from that event, such that it hits the screen
perpendicularly.  So we assign to this point here a
transverse position $x_\perp$.  An instant on the screen
corresponds to a light-front.  It corresponds to everything along the
light
ray back into the past on a light-front.  To make this more precise,
we can introduce a set of light-front
coordinates by gauge fixing the
metric of space-time to have the form
\begin{equation}
ds^2 = g_{+-} d{x^+} d{x^-} + g_{+i} d{x^+} d{x^i} + g_{ij}d{x^i}
d{x^j}
\label{metric}
\end{equation}
where the components $x^i$ refer to transverse space and
$(x^+ , x^-)$ are light-like linear
combinations of the time and $z$ coordinates. Assume that at
${x^-}=\infty$ the metric has the
flat form with $g_{+-}=1 , g_{+i}=0 , g_{ij} = \delta_{ij}$. We may
identify $ x^+ = z+t , x^- =t-z $.
The screen will be identified as the surface $x^- = \infty$.
The trajectories
\begin{equation}
x^+ = {\rm const}\ , \ \  x^i = {\rm const}
\label{nulge}
\end{equation}
are easily seen to be light-like geodesics and the surface
$x^+ = {\rm const} $ is a light-front. We will
follow the standard practice of using $x^+$ as a time coordinate when
doing light-front quantization . As we have seen it is also
the time at the screen.
So we are describing everything in the universe with light-front
quantization, except somehow we have to suppress the $x^-$
dependence (the distance along the light-front).  This will
have to be some kind of light-front quantization where
$x^-$ information is stored in a different way than we are used to
in quantum field theory.
What kind of ways are possible?  Let us go back to a single bit
stored as in figure~8. How do you distinguish the $x^-$ position
of the bit?  First, recall that in quantum
mechanics, we don't need  to code both the position and the
momentum.  We will code the momentum $p^+$ of the particle (along
with it's transverse position).  The longitudinal momentum
$p^+$ will be encoded with discrete light-front quantization
(DLCQ)[13].
Momentum comes in little individual
bits\footnote{Eventually we
will have to take the continuum limit of this picture,
but we will not discuss that problem in this lecture.
}
of size $\epsilon$, where we get to choose
$\epsilon \sim M_{\rm Planck}$, but one can make it even
smaller\footnote{
One will have to be very careful in any real
theory of this that even with this cutoff, the theory comes out to
be longitudinally boost invariant.
}
To code the momentum, we say that a particle
with a bit of minimum momentum, lights up one pixel on the
screen.  What about a particle with two units of momentum?
It lights up two pixels.  So the rule is that the information
is coded by the number of bits on the screen.
If there is another particle behind the first, one will
expect some kind of quantum mechanical lensing effect.
It is not at all obvious that we can't code everything
in this way.

We can get back to regular quantum field
theory when the information density is very low.  Imagine
going to this limit by keeping the momentum fixed in the partons
but letting the pixel size on the screen get small.
When the pixel size gets infinitely small, on
an arbitrarily small patch of the screen, one can code any momentum
one desires, and the full light-front quantization is restored.
The larger the momentum of the particle, the more bits it
has to have, and this picture simply tells you
to think in the light-front frame of particle with momentum
$p$ as having a number of constituents proportional to $p$,
with its area necessarily growing at least as fast as momentum.
We are back where we started, in
section 5 where we concluded that a particle of momentum
$p$ had to have constituents proportional to $p$, and that
the constituents were not allowed to get into each other's
way.

\vskip.3in
{\bf 9. Discussion}
\vskip.1in

What this picture is certainly telling us is that
we should not be doing quantum field theory at the Planck scale,
because we are
over counting the degrees of freedom by a tremendous amount.
And it strongly suggests that the Universe
is not even a continuum $2 + 1$
dimensional theory, because our arguments are based on
a lattice picture of the storage of bits of information.
So if we take this seriously, a proper description
of quantum gravity will be a discrete light-front theory
(DLCQ) with a discrete transverse plane of degrees of freedom,
with one bit of information per unit area of the transverse plane.

\tHooft\ asked the question -- if one does formulate a theory like
this,
that anything can be mapped to a surface along any direction,
what are the set of transformation rules
which would allow you to look at the information from different
directions?  We are pointing out that this
is just the light-front problem of implementing the
angular conditions -- how you represent
the rotations of the direction of the light-front
frame in a parton type description.  The physics of the
real three dimensional world is expected to be a representation
of the three dimensional rotation group.  We have to
represent this group in the theory defined on the 2D screen.

Ordinary bosonic string theory
has been written in a form which may
help establish the relationship between string theory and
the holographic principles described in this lecture [2,10].
Details can be found in the
original references.  These are `lattice string theories', where
the transverse plane is replaced by a discrete
lattice with spacing $l_s$, and this lattice spacing is kept {\it
fixed}
throughout. A lower longitudinal
momentum cutoff $\epsilon$ is introduced so that $p^+$ comes in
discrete units.

In the short term there are a number of areas in which progress can
be made. First a better
understanding of the concept of fixed points and their relation to
Lorentz boosts and high energy
scattering is possible in ordinary field theories like QCD [14].
Secondly, for a string theory to provide an interesting candidate for
a holographic theory, several missing ingredients have to be filled
in.
The first would be to show that the $1/N$ expansion reproduces the
bosonic string perturbation expansion. However, even if
this can be done, the theory can not be analyzed nonperturbatively
because of the tachyon instability. Therefore it is very important to
determine if the lattice model can be supersymmetrized. Assuming this
is possible, we can then attack the nonperturbative issues.

\newpage
\vskip.3in
{\bf Acknowledgments}
\vskip.1in
P.G.~would like to thank Charles Thorn for useful conversations.
This work was partly supported by Komitet Bada{\'n} Naukowych
grant KBN-2P302-031-05, NSF grant
NSF-PHY-9219345-A1 (L.S.), SSC fellowship FCFY9318 (P.G.),
and DOE grant DE-FG05-86ER-40272 (P.G.).

\vskip.3in
{\bf References}
\vskip.1in

1. G.~\tHooft, {\it Dimensional Reduction in Quantum Gravity},
Utrecht

\ \ \ \ Preprint THU-93/26, gr-qc/9310006.

2. C. Thorn, {\it Reformulating String Theory With the $1/N$
Expansion}, in

\ \ \ \ Sakharov Mem.~Lec.~in Physics, Ed. L.V. Keldysh and V.Ya. Fainberg,

\ \ \ \ Nova Science, NY 1992, hep-th/9405069. See also Phys.~Rev.~D19
(1979)639.

3. L.~Susskind, Phys.~Rev.~D49 (1994) 6606.

4. J.~D.~Bekenstein, Phys.~Rev.~D49 (1994), 1912.

5. See for example K.G.~Wilson's lectures, these proceedings.

6. R.~P.~Feynman, Third Topical Conference in High Energy Collisions  of
Hadrons,

\ \ \ \ Stoney Brook,N.Y.~Sept 1969.

7. J.~D.~Bjorken and E.~Paschos, Phys.~Rev.~185 (1969) 1975.

\ \ \ \  J.~D.~Bjorken, Intern.~Conf.~on Duality and Symmetry in Hadron
Physics,

\ \ \ \ Tel Aviv, (1971).

8. J.~Kogut and L. Susskind Phys.~Rep.~8 (1973) 75.

9. J.~Kogut and L.~Susskind, Phys Rev D9 (1974) 697, 3391.

\ \ \ \  L.N.~Lipatov, Yad.~Fiz. 20 (1974) 181, (Sov.~Nucl.~Phys.~20 (1975)
94).

\ \ \ \  G.~Alterelli and G.~Parisi, Nucl.~Phys.~B126 (1977) 298.

10. I.~Klebanov and  L.~Susskind, Nucl.~Phys.~B309 (1988) 175.

\ \ \ \ \ M.~Karliner, I~Klebanov, and L.~Susskind, Int.~J.~Mod.~Phys A3 (1988)
1981.

11. M.~Froissart, Phys.~Rev.~123 (1961) 1053.

12. C.~Misner, K.~Thorne and J.~Wheeler, {\it Gravitation},
   (1970) WH Freeman and Co.

13. A.~Casher, Phys.~Rev.~D14 (1976) 452.

\ \ \ \ \  C.~Thorn, Phys.~Rev.~D19 (1975) 94.

\ \ \ \ \ H.C.~Pauli and S.J.~Brodsky, Phys.~Rev.~D32 (1987) 1993.

14. R.~Perry and K.~Wilson, Nucl.~Phys.~B403 (1993) 587.

\end{document}